\pdfoutput=1
\documentclass{JINST}
\usepackage[caption=false]{subfig}
\usepackage{epstopdf}
\usepackage[normalem]{ulem}

\newcommand{\TTX}{Tetratex\textsuperscript{\textregistered}}
\newcommand{\Mak}{Makrolon\textsuperscript{\textregistered}}

\title{Enhancement of Light Yield and Stability of
Radio-Pure Tetraphenyl-Butadiene Based Coatings for VUV Light Detection in Cryogenic Environments}

\author{L. Baudis$^a$, G. Benato$^a$, R. Dressler$^b$, F. Piastra$^a$, I. Usoltsev$^b$ and M. Walter$^a$\thanks{Corresponding
author.}~ \\
\llap{$^a$}Physik-Institut, Universit\"{a}t Z\"{u}rich,\\
  8057 Z\"{u}rich, Switzerland\\
\llap{$^b$}Paul Scherrer Institut,\\
  5232 Villigen PSI, Switzerland\\
  E-mail: \email{Manuel.Walter@physik.uzh.ch}}

\abstract{
The detection of VUV scintillation light, e.g. in (liquid) argon detectors, commonly includes a reflector with a fluorescent coating,
converting UV photons to visible light. The light yield of these detectors depends directly on the conversion efficiency. Several
coating/reflector combinations were produced using VM2000, a specular
reflecting multi layer polymer, and \TTX\, a diffuse reflecting PTFE fabric, as reflector foils. The light yield of these coatings
was optimised and has been measured
in a dedicated liquid argon setup built at the University of Zurich. It employs a small, 1.3\,kg LAr cell viewed by a 3-inch, low
radioactivity PMT of type R11065-10 from Hamamatsu. The
cryogenic stability of these coatings was additionally studied. The
optimum reflector/coating combination was found to be \TTX\ dip coated with Tetraphenyl-butadiene with a thickness of 0.9\,mg/cm$^2$
resulting in a 3.6 times
higher light yield compared to uncoated VM2000. Its performance was stable in long term measurements, ran up to 100 days, in
liquid argon. This coated reflector was
further investigated concerning radioactive impurities found to be suitable for current and
upcoming
low-background experiments. Therefore it is used for the liquid argon veto in Phase~II of the {\sc Gerda} neutrinoless double beta decay
experiment.
}


\begin{document}

\section{Introduction}\label{sec:Intro}
Liquid argon is used as a scintillator in several present experiments such as
\textsc{Gerda}~\cite{GERDA_Inst}, ArDM~\cite{ArDMCommissioning}, DarkSide~\cite{DarkSide50_FirstRes}, DEAP~\cite{DEAP3600_DarkMatter} and
ICARUS~\cite{Amerio:2004ze} and a likely candidate for upcoming experiments like DARWIN~\cite{Baudis:2012bc} and
GLACIER~\cite{Rubbia:2010zz}.
Its scintillation light has a wavelength of 128\,nm which is below the transmittance of quartz, a commonly used window material of
light detection devices like photomultiplier tubes (PMTs). The light is usually converted to longer wavelength by a coating containing the
wavelength shifter (WLS) Tetraphenyl-butadiene (TPB).
The focus of developments presented in this paper is set on the production of square meter scale WLS reflectors showing long-term stability
in liquid argon, a high light yield and low radioactivity. The light yield was measured for coatings applied to reflective
materials. Many of these coatings are
transparent, hence they allow a direct transfer to applications where transmitted light is detected, e.g. PMTs. The coated reflector
installed as part of the liquid argon veto of \textsc{Gerda} is an outcome of the developments presented in this paper. The veto
is an important part of the upgrade to Phase~II~\cite{GerdaUpgrade}. Earlier developments for LArGe~\cite{Agostini:2015boa} and long
term stability issues observed therein were the starting point of the presented developments.

\section{Description of Coatings, Reflectors and Coating Procedure}\label{sec:coat_descr}
Coatings are applied onto VM2000 Radiant Mirror foil,
254\,$\mu$m thick \TTX\ and Copper. VM2000 is a multilayer
polyester foil with >98\,\% specular reflectance from 3M$^{TM}$~\cite{3M} (very similar or possibly identical to Vikuiti$^{TM}$ ESR). It
is stiff and available only with
glue on one side which was removed before application of the coatings. VM2000 acts as a wavelength shifter by itself emitting light of
approximately 420\,nm for an excitation wavelength of 260\,nm as shown in Fig.~\ref{fluorimeter}. \TTX\ is an expanded PTFE fabric 
from Donaldson~\cite{Donaldson} with high diffuse reflectance. It is very
flexible, can be stretched easily and resists strong
forces before ripping. The stretch is mostly
elastic making it suitable to span over uneven surfaces. Cu is of importance as it is a known low
radioactivity material commonly used in many low background experiments. It was treated with acetic acid before coating in order to
remove copper oxide.

All coatings are produced by pulling the reflectors at an angle of about
45$^\circ$ through a solution containing the respective fluorescent substance. For this purpose, a dedicated tool was developed consisting
of a pot containing the solution and two rotating cylinders. The reflector is inserted between the two 
cylinders and pulled trough the bath situated below the second cylinder.

Four coatings using \Mak, TPB and Dichloromethane with different ratios of TPB to
\Mak\ were produced. The amount of TPB + \Mak\ adds up to 31.3\,g per liter of
Dichloromethane in each solution. In such a coating the TPB is embedded into a polymer matrix which provides mechanical
stability. Additionally, pure TPB was dissolved in
Dichloromethane with concentrations from 5.8\,g/l to its saturation at 46\,g/l. A coating containing 3g/l TPB + 30 g/l
polystyrene dissolved in Toluene, which had been found to be a stable and efficient coating for liquid argon
scintillation light in~\cite{PfeifferThesis}, was provided by the MPIK Heidelberg. Commercial plastic scintillators are
also good candidates for efficient wavelength shifters. For this reason UPS-923A, BCF-10 and BC~408 were dissolved in Toluene with a
concentration of 33 g/l and coated on VM2000. The thickness of each coating was determined by measuring the weight of a small
reflector sample before and after coating. The thickness of latter coatings were 0.16 mg/cm$^2$, 0.26 mg/cm$^2$ and 0.25 mg/cm$^2$
respectively. It was measured five times for VM2000 with 80\,\% \Mak\ and
20\,\% TPB resulting in a thickness of 0.082\,mg/cm$^2$ with a variation of $\pm10\,\%$. Four coatings of VM2000 with
polystyrene\,+\,TPB
resulted in 0.073\,mg/cm$^2$ with a variation of $\pm15\,\%$.

With the coating tool described above, it is not possible to avoid an undetermined amount of coating being also
deposited on the backside of the reflector. For this reason coatings on VM2000 produced to measure the thickness were done vertically
dipping the sample into a bath and slowly pulling it out again. Pulling vertically instead of under an angle of 45$^{\circ}$ might result
in a slightly thinner coating than obtained using the coating tool.

The fabric nature of \TTX\ causes the solution to be soaked up by the foil. As a result, the coating is on both sides of
the foil as well as within
it. This allows to coat with a solution containing only TPB and the solvent. Applying this coating, \TTX\ shrinks by
$\approx$\,40\,\% and becomes curly while the solution is drying. This can be avoided by fixing it to
a PTFE sheet. The deposited amount of TPB
was in the range of 0.17 to 1.18\,mg/cm$^2$ depending on the concentration of TPB. The thickness of these coatings was reproducible within
$\pm$3\,\%. This uncertainty is assumed for the thickness of all coatings of this type.

\begin{figure}
  \subfloat{\includegraphics[width = 0.49\textwidth]{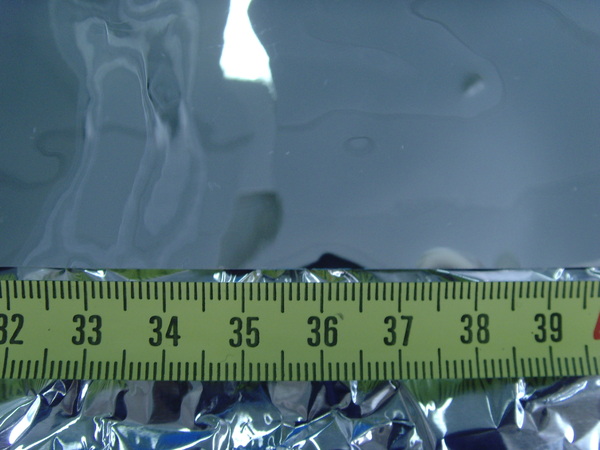}}\hfill
  \subfloat{\includegraphics[width = 0.49\textwidth]{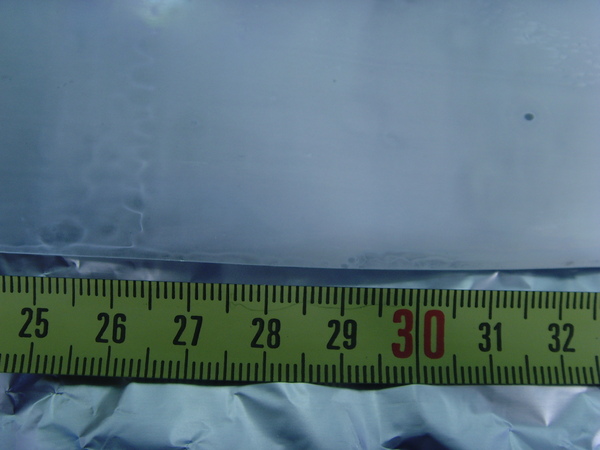}}\\
  \caption{Left: a picture of 10/1 polystyrene/TPB coated on VM2000. This coating is  completely transparent. Right: a typical
picture of a coating
with TPB\,+\,\Mak\ with up to 60\,\% of TPB is shown. These coatings are milky.}
\label{CoatPics}
\end{figure}

\begin{figure}\centering
 \subfloat{\includegraphics[width =0.49\textwidth]{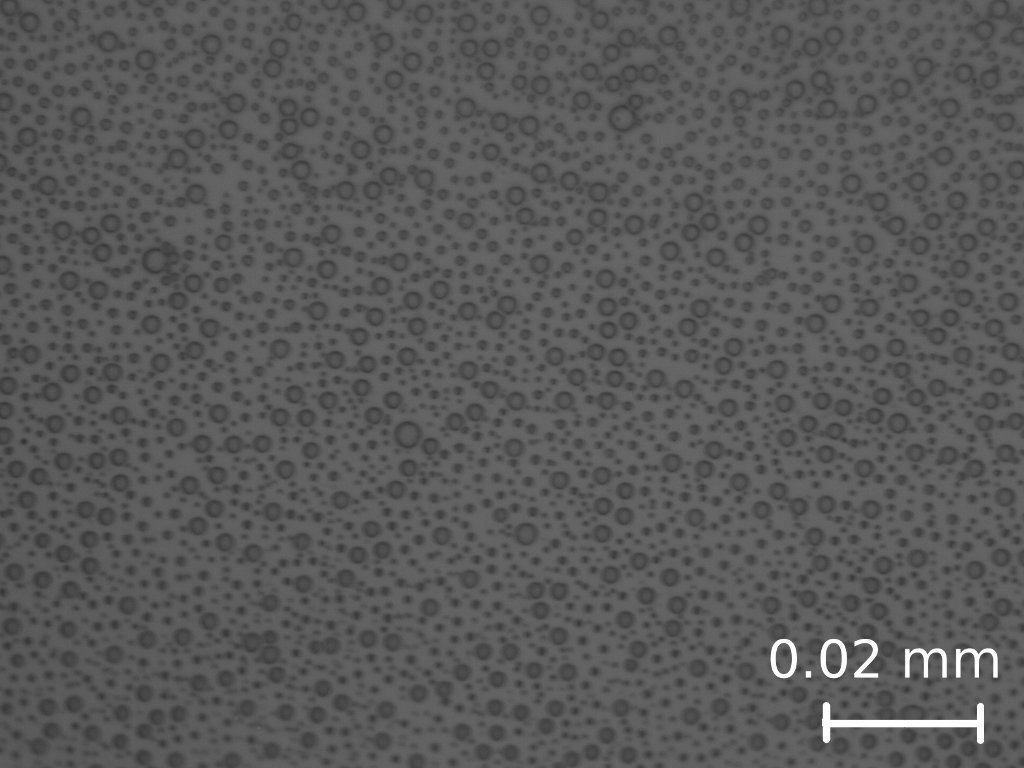}}\hfill
 \subfloat{\includegraphics[width =0.49\textwidth]{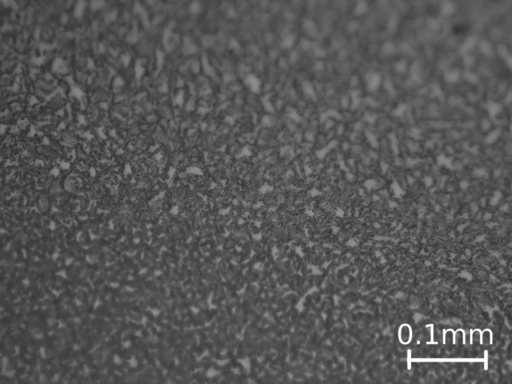}}\\
 \caption{Left: a close up for coatings with \Mak\ and up to 60\,\%TPB on VM2000 showing micro bubbles which are the
origin of the milkiness. Right: typical micro crystals forming for coatings with 80\,\% TPB and 20\,\% \Mak.}
 \label{HighTPBCoatingPic}
\end{figure}

Coated samples of all produced coating - reflector combinations were inspected by eye and with a light microscope using both UV and
visible light. polystyrene\,+\,TPB coatings are completely clear (Fig.~\ref{CoatPics} left). Under UV light they appear less bright
than \Mak\,+\,TPB
coatings, which is an indication for a worse
efficiency. Samples coated with \Mak\,+\,TPB look milky (Fig.~\ref{CoatPics} right). The higher the concentration of \Mak\
relative to TPB the more uniform the coating. The origin of the milky appearance are small bubbles of
different sizes (Fig.~\ref{HighTPBCoatingPic} left) visible under the microscope. Adding
more solvent significantly decreases the number of
bubbles per area and increases their size.
Micro crystals are forming for a concentration of 80\,\% TPB and 20\,\% \Mak\ (Fig.~\ref{HighTPBCoatingPic} right). These are most likely
crystals of pure TPB not embedded into the polymer matrix.

\begin{figure}
  \subfloat{\includegraphics[width = 0.32\textwidth]{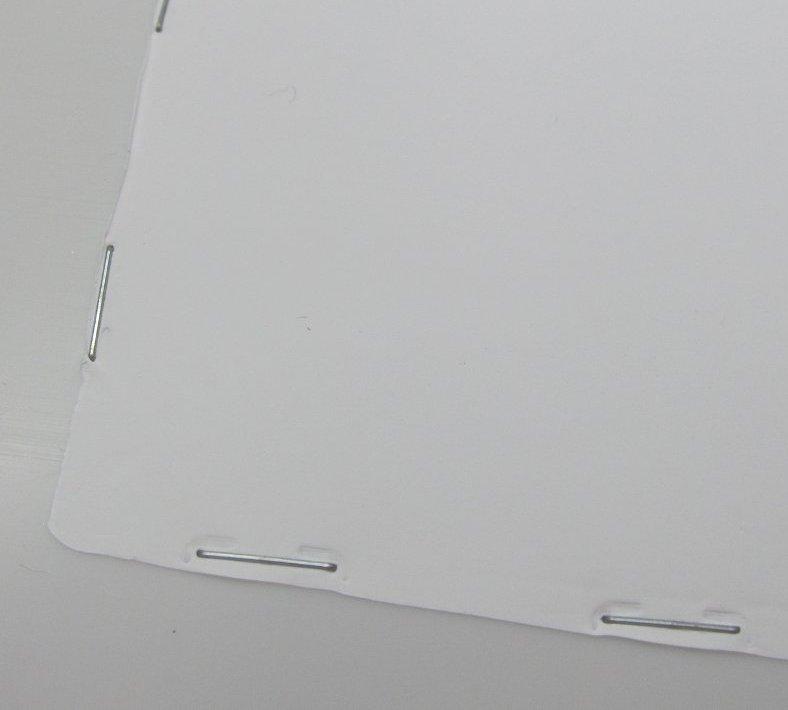}}\hfill
  \subfloat{\includegraphics[width = 0.32\textwidth]{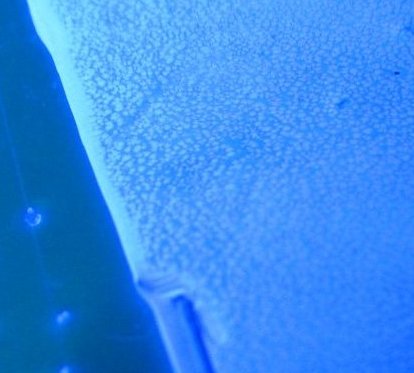}}\hfill
  \subfloat{\includegraphics[width = 0.32\textwidth]{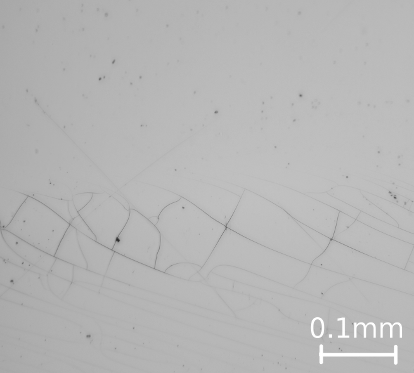}}
   \caption{Left and centre: \TTX\ coated with pure TPB and stapled onto a PTFE support. It is perfectly white under day
light (left) and shows many small spots under UV light (centre), which are
brighter than the rest. These spots have a size in the range of $\approx$\,0.1 - 1\,mm. Right: cracks visible in coatings of
TPB\,+\,polystyrene on
VM2000 after immersion into liquid nitrogen.}
\label{CoatedTTX}
\end{figure}

\TTX\ coated with pure TPB looks perfectly white under visible light, both by eye (Fig.~\ref{CoatedTTX} left) and with
the microscope. Under UV light there are small spots visible which are more bright than the rest. These spots have a size in the range of
$\approx$\,0.1 - 1\,mm. The spot size is varying even when using the same concentration of TPB in the solution. The
area covered by these spots increases with the concentration of TPB in the solution. As an example, the central picture in
Fig.~\ref{CoatedTTX} was taken
from a coating with 0.73\,mg/cm$^2$. 


Additionally, a \TTX\ and a VM2000 sample evaporatively coated with TPB as described in \cite{ArDM_WLSDevelop} 
 and \cite{Francini_VUVcarctTPB} respectively, were kindly provided by the respective groups. For the apparatus used to coat the
first sample, a capability to
coat reflector sheets with up to 120\,x\,25\,cm$^2$ is reported. For the latter a maximum size of $\approx$\,45\,x\,45\,cm is estimated from
the information provided in \cite{Francini_VUVcarctTPB}. 

In summary, the following samples are investigated:\\
\textbf{VM2000:} bare, TPB + \Mak\ in various concentrations, polystyrene + TPB 10:1, UPS-923 A, BCF-10, BC 408, pure TPB
evaporated;
\textbf{\TTX:} pure TPB dip coated in various thicknesses, pure TPB evaporated;
\textbf{Cu:} TPB + \Mak\ in various concentrations and polystyrene + TPB 10:1.

\section{Mechanical and Cryogenic Stability}
Several tests including wiping, scratching, bending and blowing air onto the sample have
been performed to gather information on the abrasive stability of all previously described coatings,
before and after immersion into liquid nitrogen. Wiping with a vinyl glove over \TTX\ coated with pure TPB results in small amounts of TPB
on the glove visible using UV light but does not leave observable traces on the foil, both under UV and visible light. The same is
true for the sample with TPB evaporatively deposited on VM2000. A summary of the mechanical stability tests together with basic sample
properties is provided in Table\,\ref{tab:stabilitySummary}.

\begin{table}
  \begin{tabular}{ l|c|c|c|c|c|c }
    Reflector	& \multicolumn{2}{c|}{ \TTX\ }	& \multicolumn{4}{c}{VM2000} 				\\ \hline 
    characteristic& 	\multicolumn{2}{c|}{elastic, fabric} 	&	\multicolumn{4}{c}{ stiff, multilayer}		\\ \hline
    coating	& dip coated	& evaporated	& evaporated	& Mak.\,$\geq$\,40\,\%	& Mak. 20\,\%	& PS 91\,\%	\\ \hline
    reflectivity&	diffuse	&	diffuse	& diffuse	&  prim. specular	& prim. diffuse & specular	\\
    bending	&	+	& 	+	&	+	&	+		&	+	&	+	\\
    blowing	& 	+	&	-	&	+	&	+		&	+	&	+	\\
    wiping	& 	o	&	-	&	o	&	+		&	-	&	+	\\
    scratching	& 	-	&	-	&	-	&	-		&	-	&	-	\\
  \end{tabular}
  \caption{Summary of coating and reflector properties as well as results from the mechanical stability tests of the most relevant samples.
All coatings contain TPB, the type and amount of admixture is given if applicable. Mak. is
used as an abbreviation for \Mak\ and PS for Polystyrene. ``+'' stands for no damage observed, ``o'' minor damage, ``-'' significant
damage.}
\label{tab:stabilitySummary}
\end{table}

Each sample had been cycled 10 times and stored in liquid nitrogen
for 72\,days. Afterwards all samples where inspected for cracks using a microscope. TPB\,+\,polystyrene coatings developed
cracks (see Fig.~\ref{CoatedTTX} right) already after the first immersion into liquid nitrogen. These cracks developed preferably at
sub-mm to mm scale locations having a thicker coating due to imperfection of the coating method. None of the other coatings developed any
cracks, not even when bending the coated reflector while immersed in liquid nitrogen. No change has been observed in the mechanical
stability after immersion into liquid nitrogen for any sample.

\section{Fluorescence Spectra at an Excitation Wavelength of 260\,nm}\label{sec:fluorimeter}

\begin{figure}
 \begin{center}
  \includegraphics[width=1\textwidth]{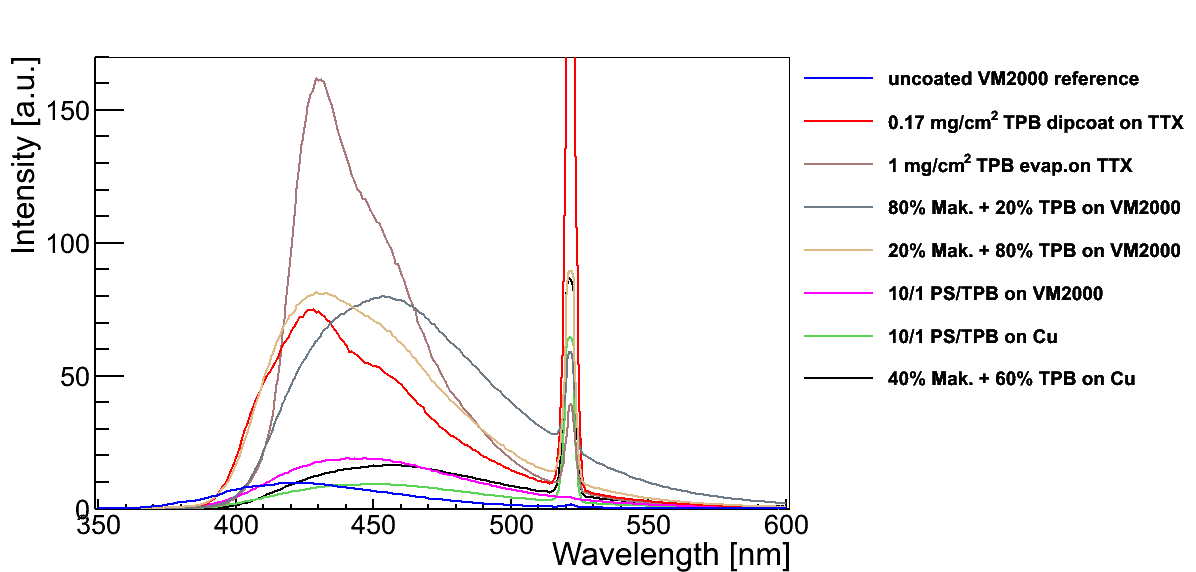}
\caption{Typical fluorescence spectra of coated reflector foils using an excitation wavelength of 260\,nm. The peak at 520\,nm originates
from 
diffuse reflected light at twice the excitation wavelength and is not related to fluorescence. TTX stands for \TTX.} 
\label{fluorimeter}
 \end{center}
\end{figure}

The emission spectra of all coatings described in Section~\ref{sec:coat_descr} except for the commercial scintillators
have been measured at an excitation wavelength of 260\,nm employing a fluorescence spectrometer of type Cary Eclipse from the Varian
company~\cite{VarianComp} at the MPIK Heidelberg. It employs a
monochromator on the excitation and emission side to select a specific wavelength from a continuous spectrum. It is however transparent for
light with wavelengths of multiples of the
selected one. The device automatically corrects for its spectral
sensitivity and is regularly calibrated by the producer. Samples were installed under an angle prohibiting specular reflectance into the
spectrum analyser. A representative
selection of these measurements is shown is Fig.~\ref{fluorimeter}. The peak at 520\,nm
($=2\cdot260$\,nm) is diffusively reflected light of twice the excitation wavelength. This peak is not related to the shifting
efficiency but corresponds to the diffusivity of the sample. A large peak at 520\,nm as observed for dip coated \TTX\ does hence mean a
large diffuse component of the reflectivity (diffuse reflector). Otherwise a small peak as for uncoated VM2000 means a small diffuse
component (specular reflector). VM2000 (blue lines in
Fig.~\ref{fluorimeter}) was used
as a reference, to monitor the time stability of the measurement apparatus and the reproducibility of the
measurements. The amplitude of this spectrum was varying by approximately $\pm$\,10\,\%.

The dark brown line in Fig.~\ref{fluorimeter} shows TPB evaporatively deposited onto \TTX. For the given thickness of
1\,mg/cm$^2$ the spectrum is expected to be strongly dominated by the properties of TPB, independent of the used
substrate (see Fig.~8 right in~\cite{ArDM_WLSDevelop}).

Coating \TTX\ by wet dipping with 0.17\,g/cm$^2$ (red line in Fig.~\ref{fluorimeter}) results in more light emitted at the low
wavelength side of the spectrum 
but a smaller overall efficiency. As shown in section~\ref{sec:VUV_Res} the light yield of this coating can be increased
by a thicker layer of TPB. 

Coatings with TPB\,+\,\Mak\ on VM2000 and \Mak\ concentrations of 40, 60 and 80\,\% have the same spectral shape and light yield
within the
uncertainty of the measurement.
Representatively shown is the spectrum for 80\,\% \Mak\ and 20\,\%
TPB (gray line in Fig.~\ref{fluorimeter}). The spectral shape of coatings with 80\,\%TPB and 20\,\% \Mak\ on VM2000 is dominated by TPB
with a small effect of the \Mak\ content.

Coatings on Cu (green and magenta lines in Fig.~\ref{fluorimeter}) have a significantly reduced light yield compared to the same
coatings
on VM2000. This shows the importance of the reflector itself. These results are however still important for detector
components which can not be covered by reflector foils.

\section{Light Yield in Gaseous and Liquid Argon}

The scintillation light of argon has a shorter wavelength of 128\,nm compared to the 260\,nm of the fluorescence spectrometer. The
light yield and emission spectrum from the reflector foil can
be different for these two wavelength which makes it necessary to confirm the result at the excitation wavelength of interest. A special
setup was built for this purpose, also to study possible effects of argon gas
and liquid argon (87\,K at standard pressure) both on the stability and light yield. The latter can be affected in two
ways. First, a change in temperature has been observed to cause a slight change in the emission spectrum of TPB~\cite{Francini_VUVcarctTPB}
and a thermal contraction of the reflector. Second, the refractive index is different for liquid argon and air
resulting in changed refraction and reflection properties at the wavelength shifter\,-\,argon and argon\,-\,light
detection device surfaces.

\subsection{Scintillation Process in Argon}
Charged particles crossing gas or liquid argon cause excitation and
ionization of argon atoms which collide with other
Ar atoms to form Ar$_2$ molecules, so called excited dimers or excimers~\cite{TimeDepLumi}. These excimers are produced
in a singlet and a triplet state with life times of 7\,ns and 1600\,ns,
respectively~\cite{ScintilPulseShape}. The decay of these molecules emits light with a wavelength of 128\,nm. Impurities such as oxygen,
water or nitrogen provide an additional non radiative decay channel by collisions~\cite{Acciarri:2009xj}. This effect is significantly
larger on the triplet state due to its longer life time. This makes the triplet life time a good measure of the purity. 

\subsection{Experimental Setup}
\begin{figure}
\begin{center}
  \includegraphics[width = 0.47\textwidth]{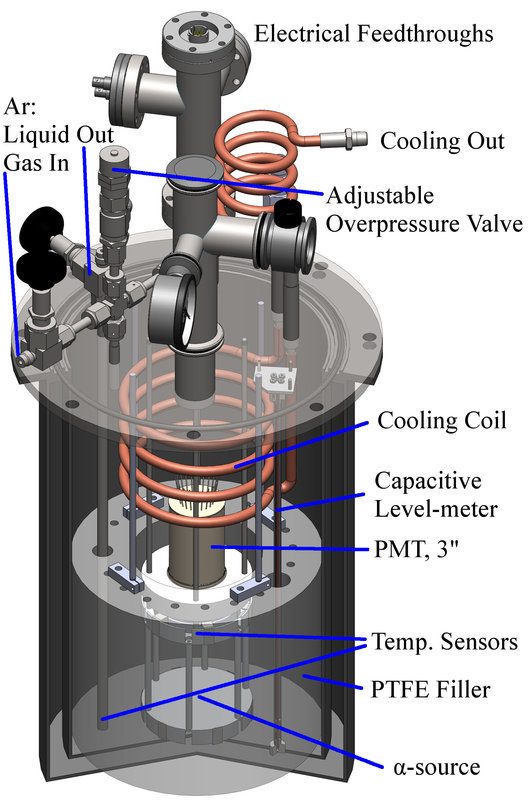}
    \includegraphics[width=0.52\textwidth]{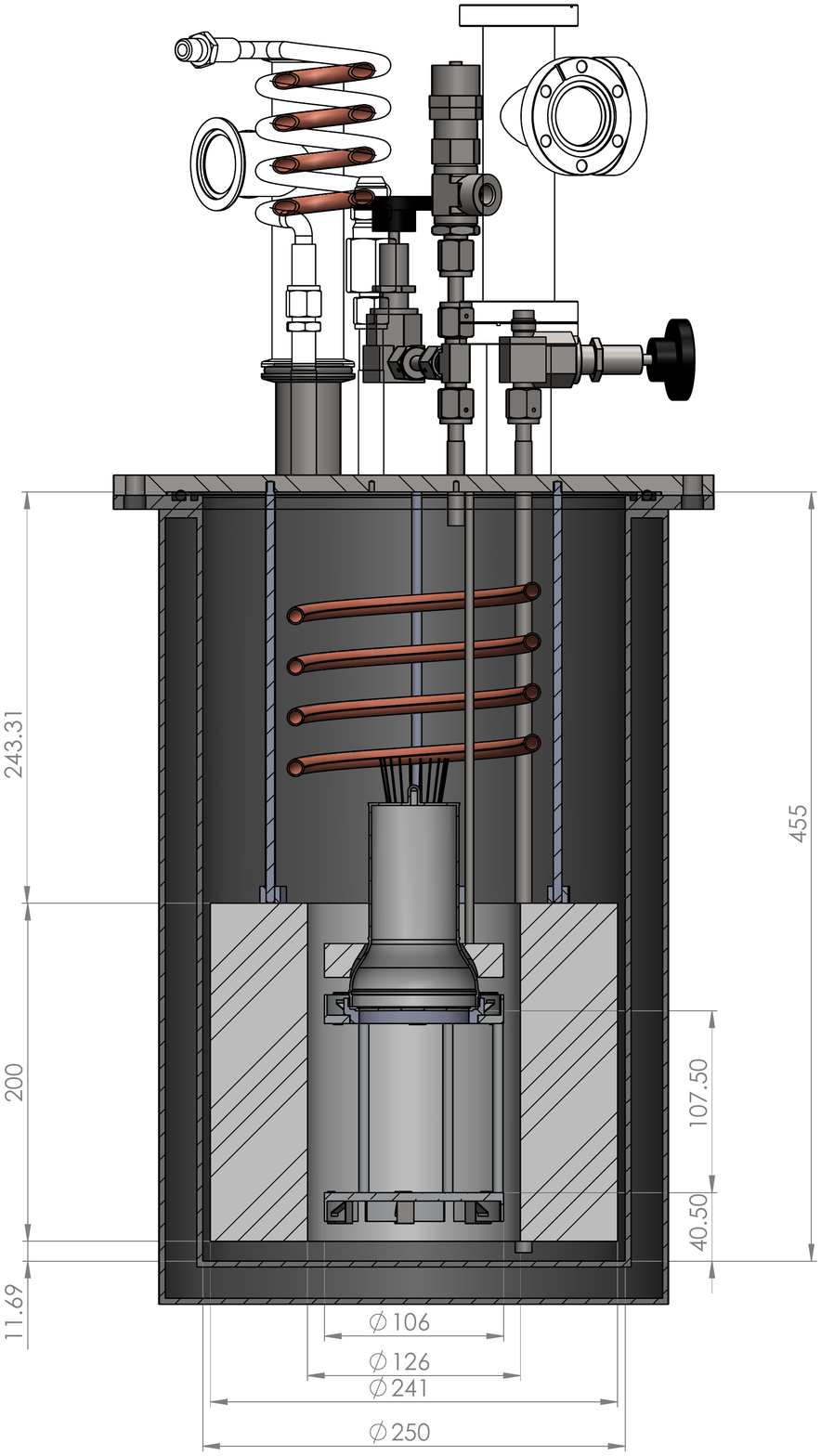}
\end{center}
  \caption{Schematic figure of the setup to measure light yield in liquid argon.} 
  \label{LArS}
\end{figure}

The liquid argon setup is shown in Fig.~\ref{LArS}. The inner volume is evacuated by a turbo molecular pump to a pressure below
$5\cdot10^{-5}$\,mbar. Argon with a purity of 99.9999\,\% (6.0) is filled from a gas bottle. Liquid nitrogen is used to
 condensate argon on a Cu cooling coil. The cooling power is regulated by the flow rate of evaporated nitrogen at the
output side of this coil. The liquid 
level is measured employing a capacitive level meter. The temperature
inside the chamber is measured by two PT-100 resistant thermometers. One is located at the bottom of the cryostat and
one just above the PMT face to additionally verify a sufficient fill level. A hollow PTFE cylinder is placed
between the cryostat wall and the active volume when operating in liquid to reduce the amount of argon needed.

A reflector foil to be measured is installed, forming the mantle of the active volume of $\approx 1.3$\,kg of liquid argon defined by
an aluminium cylinder. The cylinder consists of a top ring holding the PMT and a bottom ring holding an $^{241}$Am source. Dimensions are
provided in Fig.~\ref{LArS} right. Both rings have a rim directed outwards
and are connected by six aluminium bars equally distributed on the outer edge of the cylinder inside the active volume. The reflector foil
is wrapped around this
cylinder and clamped to the top and bottom rim cut to a length allowing 5 -- 20\,mm of overlap. It
is taken care to minimise gaps between the rim and the reflector foil as well as where the foil is overlapping. The shifted light is
detected with a 3\,inch
low radioactivity PMT of type R11065-10 from Hamamatsu~\cite{HamamatsuComp}, closing the top part of the active volume. Scintillation light
hitting the PMT directly is absorbed by its quartz glass window and not detected. The PMT signals are recorded without
amplification using a 250\,MHz ADC with 12\,bit resolution.

\begin{figure}
 \begin{center}
  \subfloat{\includegraphics[width=0.5\textwidth]{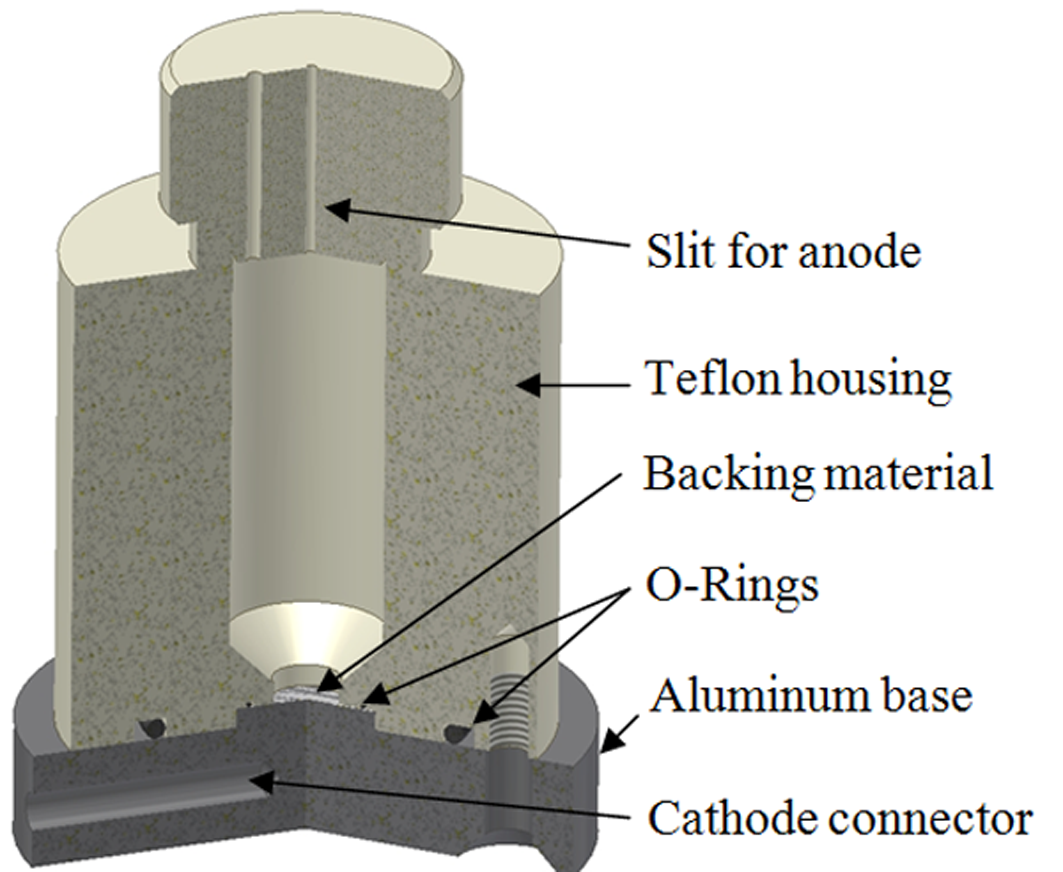}}
  \subfloat{\includegraphics[width=0.5\textwidth]{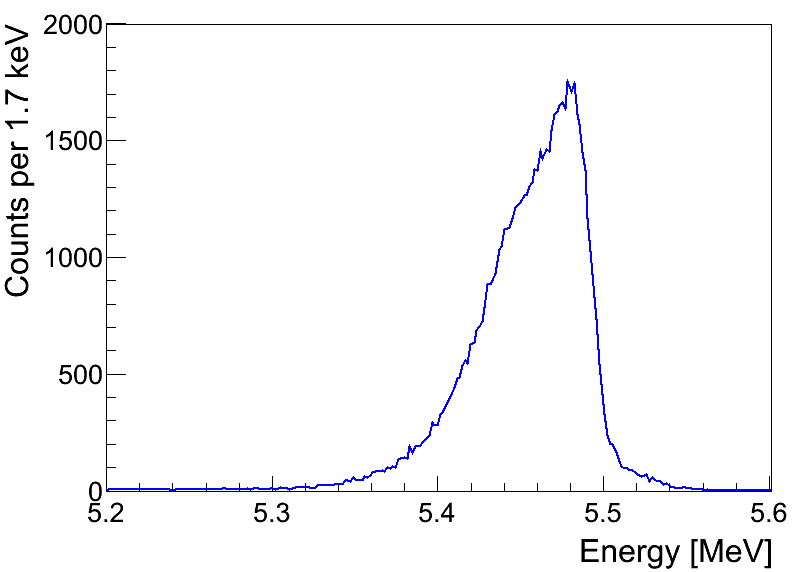}}
 \end{center}
  \caption{Left: Scheme of the cell for electrodeposition. Right: Spectrum of the final $^{241}$Am sample recorded with
a Canberra PIPS detector (active area 600\,mm$^2$) during 16.3\,h.}
  \label{ElectroDepCell}
\end{figure}

\paragraph{$^{241}$Am $\alpha$ source for cryogenic liquids}\label{sec:source}
The $^{241}$Am $\alpha$ source was prepared by means of the Molecular Plating technique \cite{MolecularPlating1},
\cite{MolecularPlating2}, \cite{Usoltsev2014297}. Electrodeposition was performed from a
$^{241}$Am nitrate solution in isopropyl alcohol using the cell shown in Fig.~\ref{ElectroDepCell}, with an
inner diameter of 16\,mm and a diameter of 7\,mm of the deposition zone. Prior to use, 25\,$\mu$m stainless
steel backing material was treated with a diamond polish (1\,$\mu$m grain size) and subsequently rinsed with distilled water
and ethanol. The $^{241}$Am solution (about 10\,ml) was filled into the cavity of the deposition cell. A spiral-shaped
Pt wire covering the required circular deposition area (0.3\,cm$^2$) served as a counter electrode and was
placed at a distance of about 1\,cm from the cathode. 1\,h electrodeposition at 600\,V allowed for production of
homogeneous 30\,Bq $^{241}$Am source with qualitative yield (i.e. 80 to 90\,\% of Am is deposited).

An effective $^{241}$Am thickness of 0.62\,ng/cm$^2$ was expected according to the measured $^{241}$Am activity, but a
visible deposition was observed pointing to a considerable high amount of inactive carrier material (zircon iodate). This carrier
was
used in preparation steps producing the $^{241}$Am stock solution to enhance the chemical yield.
Therefore, the resolution of the $\alpha$-peak in the spectrum (Fig.~\ref{ElectroDepCell}) was degraded in a way that
the
satellite $\alpha$-line of $^{241}$Am (5.388\,MeV, 5.443\,MeV, 5.511\,MeV and 5.544\,MeV) could not be separated from the
main $\alpha$-line at 5.486\,MeV.

The mechanical stability of the deposition was tested after the preparation using $\gamma$-spectroscopy. The sample was placed 2\,mm
in front of the carbon fibre window of a Canberra BEGe-2825 high purity Ge-detector and measured for 15\,min. The recorded, background
subtracted peak area of the characteristic $^{241}$Am $\gamma$-line at 59.54\,keV was 3496. The sample was immersed in liquid nitrogen for
$\approx 5$\,min and warmed up to room temperature afterwards. This procedure was repeated 10 times. Afterwards, a second $\gamma$-spectrum
was recorded under identical conditions yielding a peak area of 3532. The count rate before and after these treatments are therefore
identical within the statistical uncertainties.

\subsection{Analysis Methods}\label{sec:AnaMeth}
The light yield of each coating is determined in terms of the number of photoelectrons (pe) detected for 5.486\,MeV $\alpha$
events at a given argon purity. For this purpose, a series of analysis steps needs to be performed. In a first step
recorded waveforms are analysed offline using a modified version of an analysis software originally developed for the
ArDM~\cite{ArDMCommissioning} experiment. Relevant calculated parameters for each event are
baseline, integrals, component ratio (see below), pedestal standard deviation and an
estimate on the triplet life time.

\begin{figure}
\begin{center}
  \includegraphics[width = 0.5\textwidth]{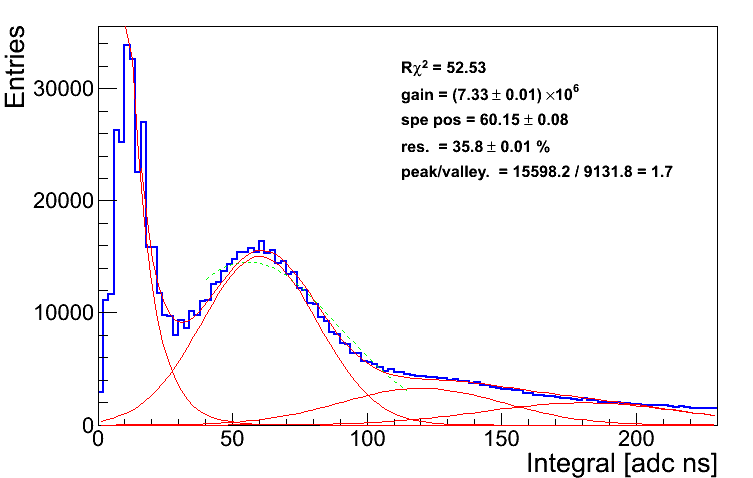}
\end{center}
  \caption{Single photoelectron spectrum at a PMT voltage of 1500\,V obtained from 10000 liquid argon scintillation pulses using a peak
finding algorithm.} 
  \label{fig:SPECalib}
\end{figure}

The PMT gain is calibrated using single photoelectron spectra. An example is shown in Fig.~\ref{fig:SPECalib}. It is obtained
directly from
light yield measurement traces using the following peak finding algorithm. When the recorded trace
exceeds 2 standard
deviations of the pedestal it is classified as a peak. The integral of a peak is calculated summing the samples above threshold
plus one sample before and after. This method results in high statistics if the argon purity is good due to many
single photo electron events present in the long tail of the triplet component.

\begin{figure}
  \begin{center}
    \includegraphics[width=0.5\textwidth]{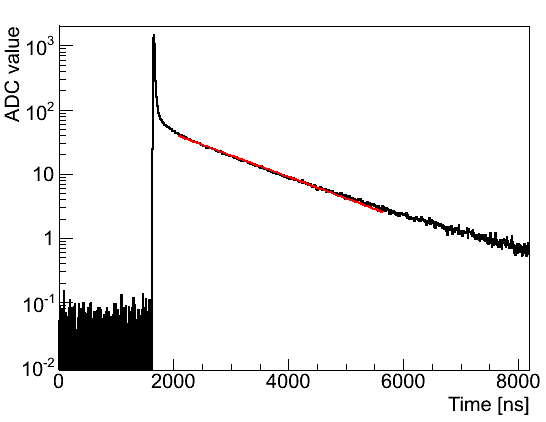}\hfill
    \includegraphics[width=0.493\textwidth]{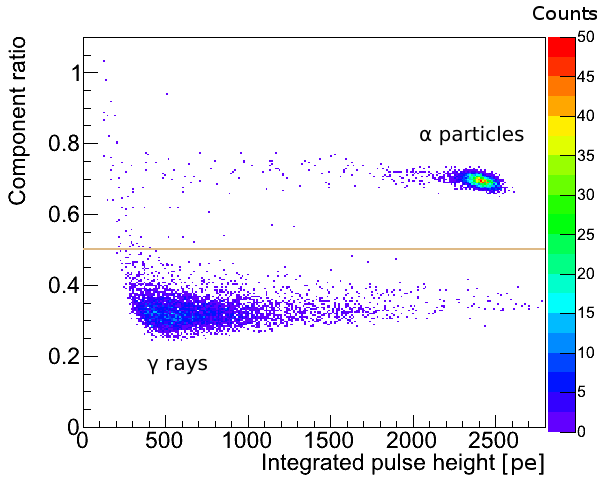}
  \end{center}
  \caption{Left: Mean trace from argon scintillation light with an exponential fit to the triplet component (here: 1305\,ns decay time).
The fast singlet and slow triplet component are clearly visible. Right: Component ratio versus integral. $\alpha$ events have a larger
component
ratio than electron recoils from $\gamma$ events, resulting in a perfect separation of background and signal. The chosen cut value is shown
by the light brown line.}
\label{CRvsIPH}
\end{figure}

The life time of the triplet state and hence the purity of the argon is determined fitting an exponential function to a
mean trace using a maximum likelihood method. This is shown in Fig.~\ref{CRvsIPH} on the left. The mean trace is calculated from 1000
consecutive events removing events with an integral of
less than 50\,pe. The systematic uncertainty determined using different fit ranges is 20\,ns. Additionally an event by event fit is
performed
to monitor sudden changes in the purity. The latter has large uncertainties and is not used for the determination of
efficiencies.

The component ratio is defined as the ratio of light in the fast component divided by the total amount of light. This ratio is smaller for
light originating from electron traces compared to nucleus traces
due to different ionisation densities~\cite{Hitachi:1983zz}. $\alpha$ events at this energy can be selected without background $\gamma$
events, producing electron recoils, due to this property of argon. This is shown in Fig.~\ref{CRvsIPH} on the right. Background events
are expected from natural contaminations of the setup materials and surrounding lab. These are visible
because of the low activity of the $^{241}$Am source of $\approx 30$\,Bq.

Maintaining a constant argon purity for all measurements and over the course of longer measurements is very difficult. For this
reason, uncoated VM2000 was measured as a reference at several argon purities. The obtained triplet life time dependent
light yield was fit with a linear function. The light yield of any other coating measured at a given triplet life time was
compared to VM2000
at the same argon purity using the fitted function.
This relative light yield $L_r$ is in good approximation a purity independent measure of the light yield which is
explained in detail in the following including the influence of wavelength shifting by impurities~\cite{Heindl:2010zz}. The rate of
photons emitted by decaying excimers $N_{\gamma}$ can be approximated by the sum of two exponential decays, 
\begin{equation}
N_{\gamma} =  A e^{-\frac{t}{\tau_1}} + B e^{-\frac{t}{\tau_2}}.
\end{equation}
The first term represents the singlet and the second the triplet state with decay times $\tau_1$ and $\tau_2$
respectively. The height of the emitted light pulse at t=0 is independent of the purity in contrary to the integral with a value of $A \cdot
\tau_1 + B \cdot \tau_2$. The decay time of non-radiative deexcitations by impurities is much larger than the
singlet life time but smaller than the triplet life time for the presented measurements. The total amount of light is hence proportional to
the measured triplet decay time $\tau_2$ plus a constant assuming no wavelength shifting by impurities. This results
in the amount of detected light $L_d$ of
\begin{equation}
  L_d(WLS) = (A \tau_1 + B \tau_2) \cdot \epsilon_{WLS},
\end{equation}\label{func:LightYieldWls}
with $\epsilon_{WLS}$ an effective efficiency containing the conversion, light collection and PMT sensitivity. The
relative light yield $L_r$ of a wavelength shifter is then the ratio:
\begin{equation}
  L_r = \frac{L_D(WLS)}{L_D(VM2000)} = \frac{\epsilon_{WLS}}{\epsilon_{VM2000}}
\end{equation}
and hence independent of the purity (the measured alpha peak position at a measured $\tau_2$ is compared to VM2000 extrapolated
to the same $\tau_2$). 

If there are impurities which shift light additionally to the coated reflector foil one has to extend the formula for $L_d$ with a
term for light absorbed by impurities. If the absorption occurs with a probability $P_i$ and detected with an effective
efficiency $\epsilon_i$ containing the conversion and collection efficiency as well as the PMT sensitivity,
Equation~\ref{func:LightYieldWls} becomes:
\begin{equation}
  L_d(WLS) = (A \tau_1 + B \tau_2) ( P_i \cdot \epsilon_i + (1 - P_i) \epsilon_{WLS}).
\end{equation}
$P_i$ depends on the purity and hence on the measure $\tau_2$ which would result in a non-linear dependency of the alpha peak
position as
function of $\tau_2$. The influence on the relative light yield can be studied by the functional form of $L_r$
in the presence of light shifting impurities:
\begin{equation}
  L_r(WLS) = \frac{P_i \epsilon_i + (1 - P_i) \epsilon_{WLS}}{P_i \epsilon_i + (1 - P_i) \epsilon_{VM200}}.
\end{equation}
The formula is valid under the assumption that each measurement is dominated by the same impurities. The assumption is
justified as our primary source of impurities is air emanating
from the pores of PTFE detector components, mainly the PTFE filler. For $P_i \rightarrow 0$ the ratio becomes equal to $
\epsilon_{WLS}/\epsilon_{VM2000}$ and equal to 1 for $P_i \rightarrow 1$ (all light shifted by impurities). A higher impurity
concentration would hence result in an underestimation of the light yield of wavelength shifter with a higher efficiency than the reference
uncoated VM2000, which is the case for all presented samples.

\subsection{Measured Relative Light Yield}\label{sec:VUV_Res}
\begin{figure}
  \includegraphics[width=0.55\textwidth]{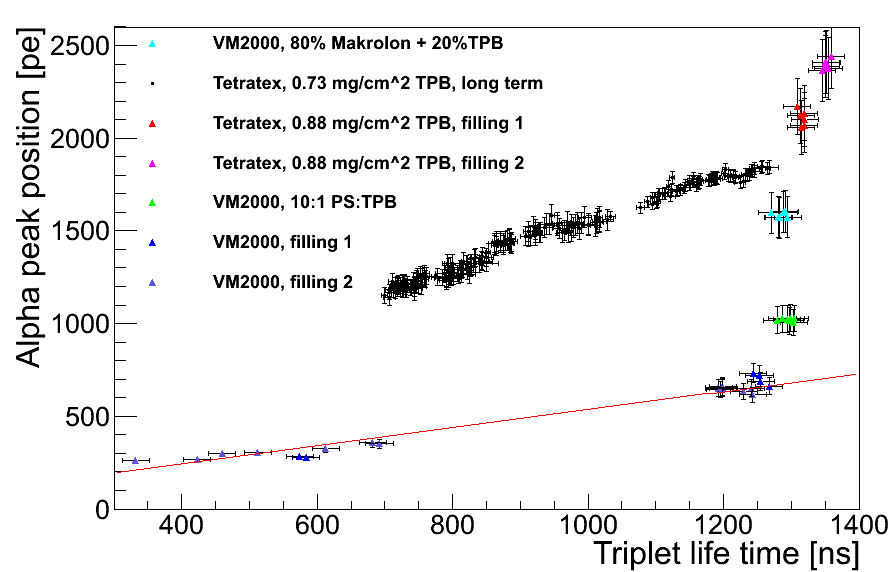}
  \includegraphics[width=0.44\textwidth]{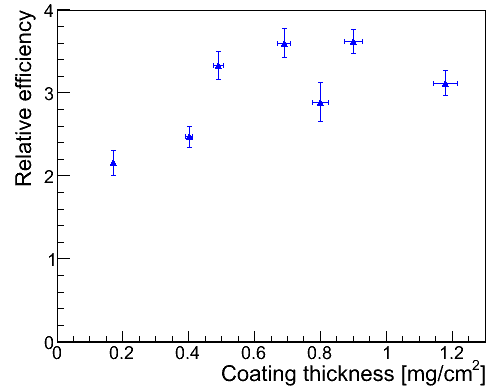}
\caption{Left: Alpha peak position of a selection of different coatings on VM2000 and \TTX\ measured in liquid argon. Error bars are
described in Section~{\protect\ref{sec:VUV_Res}}. Right: Relative light yield of \TTX\ coated with pure TPB for
different coating thicknesses.}   \label{fig:selectedEffMeasures}
\end{figure}

The light yield in the $\alpha$ peak as a function of triplet life time of a selection of coating/reflector combinations measured in liquid
argon is shown in Fig.~\ref{fig:selectedEffMeasures} on
the left. Further measurement series were performed in liquid and gaseous argon. The resulting relative efficiencies of all measured samples
are summarized in Table~\ref{tab:coatingEff}. 
\begin{table}
  \begin{tabular}{l|c c c c c c c}
    Coating	& TPB on TTX 		& Mak+TPB	& PS+TPB 	& BC408 		& BCF-10		& UPS-923A 	\\
\hline
    GAr	& 3.62\,$\pm$\,0.14	&		& 		& 1.2\,$\pm$\,0.08 	& 1.5\,$\pm$\,0.11	& 1.2\,$\pm$\,0.08
\\
    $\tau_2$(G) & 2564 -- 2981	& 		& 		& 1946 -- 2270		& 1543 -- 1991		& 2323 -- 2501	\\
    LAr		& 3.15\,$\pm$\,0.16	& 2.3\,$\pm$\,0.16& 1.5\,$\pm$\,0.11 & 		&			&	\\
    $\tau_2$(L) & 	1279 -- 1319	& 1271 -- 1295	& 1279 -- 1305	& 			& 			& 	\\
  \end{tabular}
  \caption{Light yield measured with different coatings relative to uncoated VM2000.
Coatings were applied to VM2000 except pure TPB, which has been applied to \TTX\ (TTX) by dip coating. The given relative light
yield of this coating is for the optimal thickness of 0.9\,mg/cm$^2$.
    Mak+TPB stands for 80\,\% \Mak\ and 20\,\% TPB, polystyrene+TPB for a ration of 10:1 polystyrene:TPB, G and L
for gas and liquid respectively and $\tau_2$ for the triplet life time observed in the respective measurements. For coatings measured
several times the average relative light yield is given.}
  \label{tab:coatingEff}
\end{table}

Among these measurements, \TTX\ coated with pure TPB shows the highest
light yield in addition to the previously mentioned advantages. It was further optimised by varying the coating thickness from
0.17\,mg/cm$^2$
to 1.2\,mg/cm$^2$. The latter
corresponds to the saturation concentration of TPB in Dichloromethane of $\approx$\,4.6\,g/100\,ml. The relative light yield as a
function
of coating thickness is shown in Fig.~\ref{fig:selectedEffMeasures} on the right. According to these measurements the optimal thickness is
0.9\,mg/cm$^2$. 

 The observed linear dependency of the light yield with the triplet life time is consistent with the assumption that the amount of
light shifted by impurities is small compared to light shifted by the coated reflector foils. This is confirmed by a measurement
performed without wavelength shifter which resulted in an upper limit of $< 2$\,\% of light detected compared to uncoated VM2000. The
primary source of impurities are expected
to be initial contaminations in the argon bottle (6.0) and air emanating from pores of PTFE detector components. An estimation of the 
concentration is performed employing the results of~\cite{Acciarri:2008kx} and~\cite{Acciarri:2008kv}. Therein the triplet life time was
measured as a function of O$_2$ and N$_2$ contamination. Their purest argon had a contamination of $\approx 4\cdot 10^{-3}$\,ppm O$_2$
and $\le0.5$\,ppm N$_2$ resulting in a triplet life time of $\approx 1200$\,ns which is shorter than in all liquid light yield measurements
presented in Table~\ref{tab:coatingEff}. 0.5\,ppm N$_2$ corresponds to $\approx 0.14$\,ppm O$_2$ under the assumption of air which resulted
in a decrease by $\approx 150$\,ns in their measurements. This is hence a conservative upper limit on the contaminations in liquid
measurements in the present work. The total air contamination in the gas measurements presented here is in the range of approximately 0.01
-- 0.5\,ppm as estimated employing the results of~\cite{Amsler:2007gs}.

The uncertainty on the relative light yield was found to be 7\,\%. It is calculated from
fluctuations within one measurement and fluctuations between measurements of the same coating taken under seemingly same
conditions. This includes effects originating from mounting and dismounting like the curvature of the mounted foils and small slits. The
curvature influences the solid angle of
scintillation light hitting the foil and the solid angle of illuminated reflector foil seen by the PMT. For coatings measured
several times, the uncertainty was divided by the square root of the number of measurements. 

\begin{figure}\centering
  \includegraphics[width=0.5\textwidth]{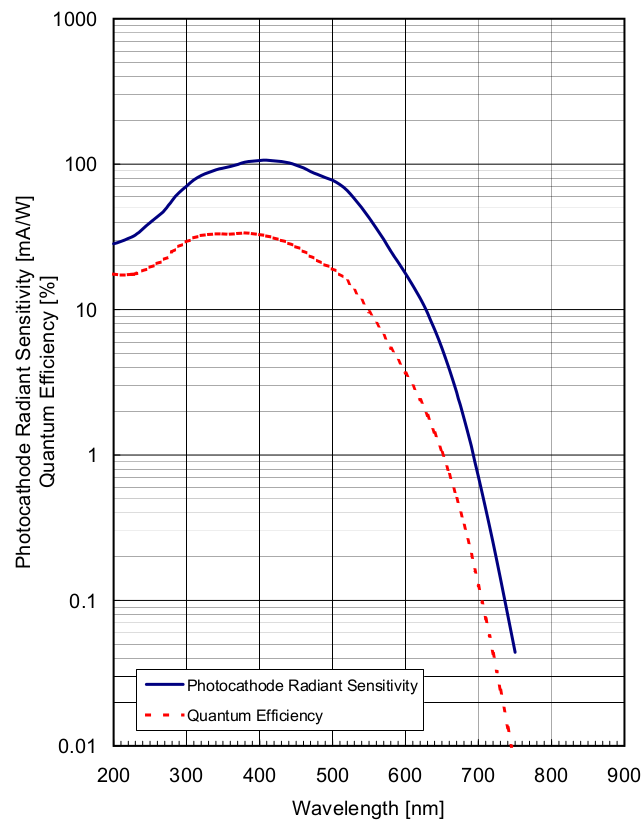}
  \caption{Photo cathode quantum efficiency and radiant sensitivity as a function of the incident wavelength at room temperature.
Below 200\,nm it decreases steeply according to the manufacturer. Radiant sensitivity is the cathode current divided by the incident
radiation flux in watt. Figure provided by Hamamatsu~\cite{HamamatsuComp}.}
  \label{fig:pmtQe}
\end{figure}

For the results presented in Table~\ref{tab:coatingEff} a PMT of type R11065-10 was employed. In order to transfer these results to
other light detection devices it is essential to know the spectral response of this individual PMT which is provided in Fig.~\ref{fig:pmtQe}
as measured by Hamamatsu~\cite{HamamatsuComp}.

\subsection{Long-term Stability in Liquid Argon}
\begin{figure}\centering
\includegraphics[width=0.8\textwidth]{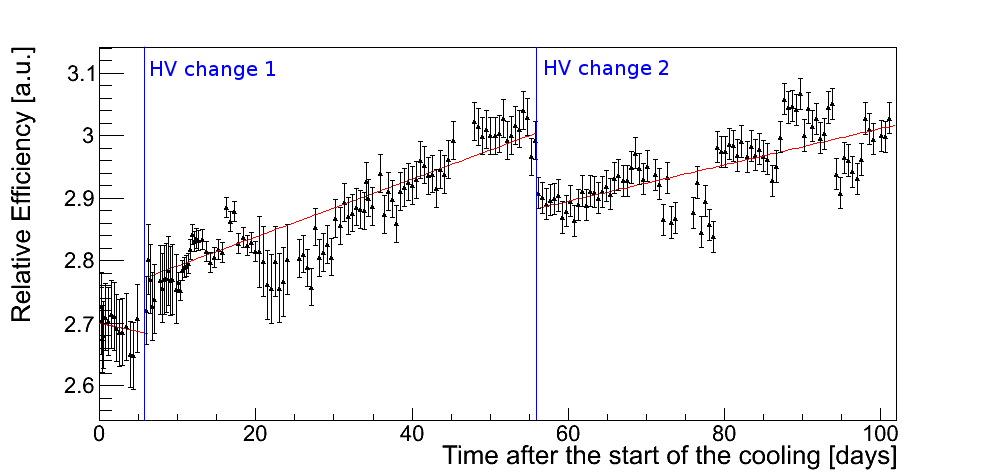}
\caption{Relative light yield of \TTX\ dip-coated with TPB as a function
of time in liquid argon, measured relative to uncoated VM2000. Time periods
with 1500\,V, 1475\,V, 1490\,V PMT voltages were fitted separately with a linear
function. Error bars as described in Section~{\protect \ref{sec:VUV_Res}}.}
\label{fig:longTermStab}
\end{figure}

The long-term stability of \TTX\ dip coated with pure TPB was tested with a continuous measurement of 100 days in liquid argon.
The purity was
decreasing within this period from an initial triplet life time of $(1260\pm20)$\,ns to $(700\pm20)$\,ns. The initial life time
corresponds to a maximum contamination of $\approx 0.5$\,ppm N$_2$ and $\approx 0.14$\,ppm O$_2$ as described in the previous section. The
final triplet life time of 700\,ns corresponds accordingly to $\lesssim 1.2$ of O$_2$ and $\lesssim 4.4$ of N$_2$. The relative light yield
is expected to be insensitive to changes in purity if the amount of light shifted by impurities is negligible compared to the coated
reflector. Otherwise it would decrease with increasing impurity concentration as described in Section~\ref{sec:AnaMeth}.
The PMT
voltage had to be changed once due to sparking in the chamber or PMT and once 
to obtain a better separation of the single photo electron peak from noise. An anti-correlation was observed between gain
and
light yield fluctuations which shows that the gain change from one to the next run is small compared to uncertainties in
the gain determination procedure. For this reason, a mean gain
is used for periods without clear gain changes. An apparent increase in light yield of
 (13.8$\pm$7)\,\% was observed over the period of 100 days (see Fig.~\ref{fig:longTermStab}) excluding impurities to have a
significant contribution to the shifted light. Error bars reflect the uncertainty on the gain
(including systematics) and on
the extrapolation for equal triplet life times. The gains of the measurements in the clusters around day 75 and day 95 which 
show a light yield lower than compatible with the red curve could not be determined and were taken as the average gain in that
region. The decrease might hence not be caused by a decrease in light yield but in gain. The voltage changes are marked
in the plot. There is no clear reason for fluctuations at other times. The light yield in gas argon measured before and after the
long-term measurement
(in liquid) shows an increase of (3.9$\pm$\,7)\%. This verifies the stability of the coating and is
 consistent with both a constant and an increasing light yield.

The scintillation efficiency is dependent on the electric field \cite{fieldDepLY}. In particular, a reduction in the electric field strength
above the $\alpha$ source due to charge accumulation, e.g. at the PMT window, reflector foil or PTFE
holder could result in an increased amount of scintillation light. The origin of the electric field in the setup is the voltage
of the PMT photocathode relative to the grounded metal parts of the chamber. It was simulated using Comsol
Multiphysics~\cite{Comsol}. The field strength 28\,$\mu$m above the source was found to be 1400\,V/m without accumulated charge. This low 
field strength is expected to have no observable effect with the given setup and hence disfavours charge accumulation as the primary effect
for the apparent light yield increase.

\subsection{Stability to Air Exposure}
The relative light yield of a \TTX\ sample coated with 0.88\,mg/cm$^2$ of TPB was measured in liquid argon to be 3.0$\pm$0.2. After
this
measurement is was stored in air for 51 day. The sample was protected from incident light by an aluminium foil but exposed to humidity and
oxygen from the air. This
sample was reinstalled and the relative light yield was measured again. It was found to be 3.3$\pm$0.2, which is consistent within
the
uncertainties with the previous measurement. Wipe tests and inspections of the foil and setup with a UV lamp showed no signs for mechanical
degradation of the coating. This confirms that the light yield and mechanical stability is not affected by air and ambient humidity.

\section{Radiopurity} 
Traces of radioactive isotopes in \TTX\ and TPB have been identified using three complementary techniques: $\gamma$ ray spectrometry,
inductively coupled plasma mass
spectrometry (ICP-MS) and radon emanation. 

The $\gamma$ ray spectrometry was performed by the UZH high purity Ge detector facility, Gator~\cite{GatorSetup}.
It is operated underground at Laboratori Nazionali del Gran Sasso (LNGS) of INFN. Thanks to the ultra-low background
($\sim100\,$counts/day in the
[50--2700]\,keV energy range) it is one of the world's most sensitive germanium spectrometers. $\gamma$-ray spectrometry is the only non
destructive technique sensitive to all primordial (e.g. $^{40}$K and
isotopes belonging to the $^{238}$U and $^{232}$Th decay chains), and anthropogenic radio-isotopes (mainly $^{137}$Cs).

The activities of the radio-isotopes are determined from the intensity of their most prominent $\gamma$ lines (e.g. the lines with the 
highest branching ratios) as described in~\cite{GatorSetup}.
This technique allows the detection of a break in the secular equilibrium of a radioactive chain.
The photo-absorption efficiency of each $\gamma$ line is calculated by means of Monte Carlo simulations performed with a detailed
GEANT4.9.3~\cite{Agostinelli2003250}
model of the sample and detector.
In order to simulate properly each isotope or decay chain the G4RadioactiveDecay class~\cite{G4RadioactiveDecay} of the GEANT4 package is
used, where all the branching ratios of different $\gamma$ lines are taken into account.

A 208\,g TPB coated \TTX\ foil was measured for 10.5 days in the Gator cavity.
Before the measurement the foil was folded to a rectangular shape with dimensions $10\times15\,$cm$^{2}$ (comparable to the Ge crystal) and
an average thickness of $4.5\,$cm. 
The sample was placed on top of the detector's cryostat a few mm from the sensitive crystal.

\begin{figure}
  \includegraphics[width=\textwidth]{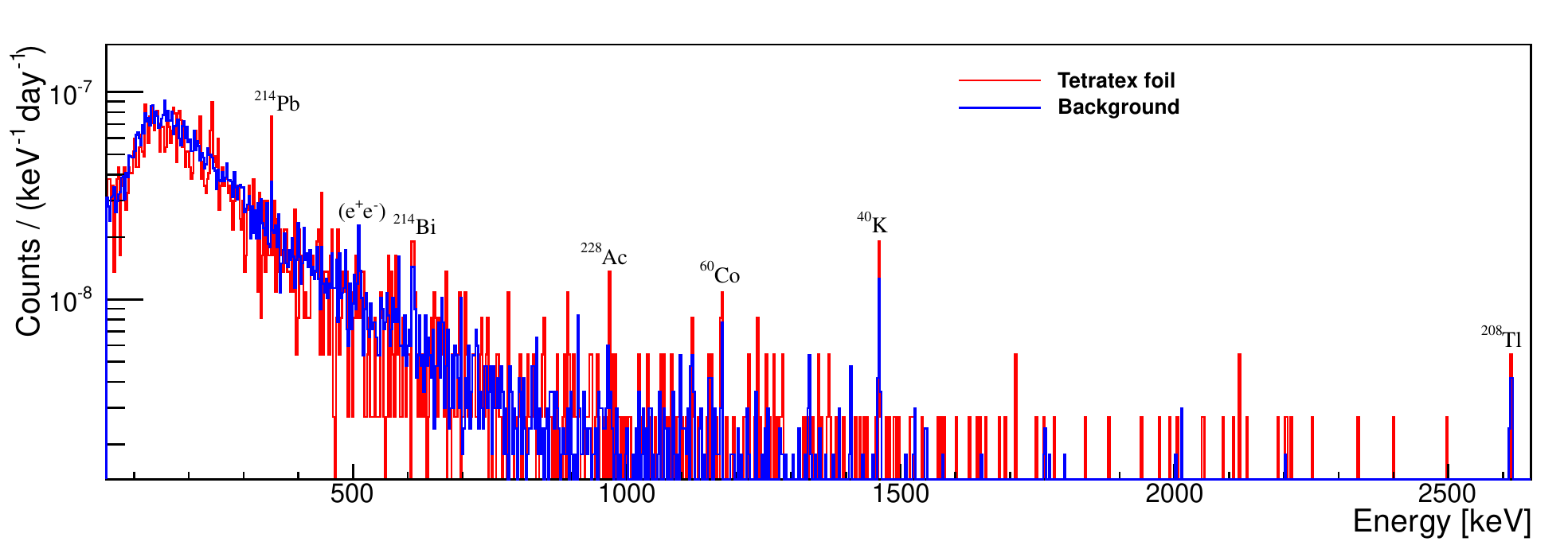}
  \caption{Spectral count rate of the TPB coated \TTX\ sample (red) superimposed by the Gator background (blue).}
  \label{fig:GatorSpect}
\end{figure}

The spectrum of the coated \TTX\ is shown in Fig.~\ref{fig:GatorSpect} (red) and compared to the background (blue) acquired with a 47.8
days run. For all isotopes of interest the rates of the $\gamma$ lines are compatible with background expectations (a detection
claim is based on a statistical test with 5\% significance level). In Table~\ref{tab:TTXRadiopurity} the upper limits at 90\% C.L. are
reported for each isotope analysed with this technique.

ICP-MS measurements were performed by the chemical service of LNGS.
A sample of uncoated \TTX, a sample of scintillation grade TPB from Sigma Aldrich and a sample of \TTX\ dip coated with TPB
dissolved in p.a. grade Dichloromethane were analysed. The blank subtracted concentrations of $^{232}$Th, $^{238}$U and $^{40}$K
in \TTX/TPB/coated \TTX\ are 0.053/0.016/0.173\,ppb, 0.071/0.014/0.123\,ppb and 287/137/746\,ppb respectively. These values
converted into activities are given in Table~\ref{tab:TTXRadiopurity}.

\begin{table}
  \begin{tabular}{ l | c  c  c  c  c  c  c  c  c }
  [mBq/kg]	& $^{238}$U	& $^{226}$Ra	& $^{228}$Ra	& $^{228}$Th	& $^{235}$U	&
$^{40}$K	& $^{60}$Co	& $^{137}$Cs	& $^{232}$Th	\\ \hline
  Gator		& < 115		& < 11.6	& < 15.3	& < 9.79	& < 6.68	& <
54.3		& < 3.43	& < 3.58	&		\\
  ICP-MS	& 1.5		&		&		&		&		&
23    		&		&		& 0.70		\\
  \end{tabular}
  \caption{Summary of radioactive traces in \TTX\ dip coated with TPB. Upper limits at
90\,\% C. L.. Uncertainties on values from ICP-MS measurements are 30\,\%. $^{238}$U, $^{226}$Ra, $^{228}$Ra,
$^{228}$Th and $^{235}$U represent the sub decay chains $^{238}$U-$^{230}$Th, $^{226}$Ra-$^{206}$Pb,
$^{228}$Ra-$^{228}$Ac, $^{228}$Th-$^{208}$Pb and $^{235}$U-$^{207}$Pb respectively.}
\label{tab:TTXRadiopurity}
\end{table}

The emanation rate of $^{222}$Rn was measured by the Max-Planck-Institut f\"ur Kernphysik in Heidelberg. 373\,g of uncoated \TTX\ was loaded
into a vacuum chamber, pumped and flushed with Rn depleted He. After this cleaning procedure emanating $^{222}$Rn is accumulated in the
vacuum chamber and transferred to proportional counters employing Rn depleted He gas. The number of Rn decays is detected in the
proportional counters and converted into a Rn emanation rate. An upper limit of 54\,$\mu$Bq/kg was found at 90\,\% C. L.. A more detailed
description of the
setup and measurement procedure can be found in \cite{RnEmanationSetupMPIK}. 

\section{Conclusion}
The light yield of several reflector/coating combinations have been measured in gaseous and liquid argon and with a
fluorescence spectrometer. Amongst the measured samples, \TTX\ dip coated with TPB is the superior coated reflector foil. The
optimal thickness of this coating was found to be 0.9\,mg/cm$^2$ resulting in a 3.6 times
higher light yield compared to uncoated VM2000. It is stable for long term operation in liquid argon and insensitive to exposure to
ambient humidity and oxygen. The effort to produce square meter scale reflectors by dip coating is significantly smaller compared to
evaporation, thus making it a well suited solution for large experiments. The mechanical stability of the foil permits easy handling
during installation and long-term operation. It has been installed in the liquid argon veto of {\sc Gerda} Phase~II. A coating
containing 80\,\%\Mak\ and 20\,\% TPB is a good alternative if transparency is required. It has a 2.3 times higher light yield compared to
uncoated VM2000.

\acknowledgments
This work is financially supported by the Swiss National Science Foundation (SNF grant Nb. 200020-149256 and 20AS21-136660) and ITN
Invisibles (Marie Curie Actions, PITN- GA-2011-289442). We would like to thank M. Heisel, A. Smolnikov and A. Wegmann from the
Max Planck Institut f\"ur Kernphysik in Heidelberg for their technical support and discussions, H. Simgen who
performed the radon emanation measurement and Maria Laura Di Vacri and Stefano Nisi from the
LNGS chemistry laboratory for the IPC-MS measurements. We also thank for the kind possibility to use
the clean room facility ``First Lab'' at ETH Zurich, the fluorescence spectrometer at Max Planck Institut f\"ur Kernphysik in Heidelberg and
the optical microscope of the group of Hans-Werner Fink at the University of Zurich.

\bibliographystyle{ieeetr}%
\bibliography{fullBib}
\end{document}